\documentstyle[twocolumn,prl,aps]{revtex}
\begin{document}
\tighten
\title{Metal-insulator transition in a multilayer system
with a strong magnetic field}

\author{X. R. Wang and C. Y. Wong$^{*}$} 
\address{Physics Department,
The Hong Kong University of Science and Technology,
Clear Water Bay,
Hong Kong
}
\author{X. C. Xie}
\address{Physics Department,
Oklahoma State University, Stillwater OK 74078 
}
\date{\today}
\maketitle
\begin{abstract}
We study the Anderson localization in a weakly coupled 
multilayer system with a strong magnetic field perpendicular 
to the layers. The phase diagram of 1/3 flux quanta per  
plaquette is obtained. The phase diagram shows that a 
three-dimensional quantum Hall effect phase exists 
for a weak on-site disorder. For intermediate 
disorder, the system has insulating and normal metallic 
phases separated by a mobility edge. At an even larger 
disorder, all states are localized and the system is 
an insulator. The critical exponent of the localization 
length is found to be $\nu=1.57\pm0.10$.
\end{abstract}

There has been an increasing interest in understanding the 
localization problem in three-dimensional (3D) or multilayer 
disordered electron systems in a strong magnetic 
field\cite{Stomer,Chalker,Balents,Ohtsuki,Druist,Hill,Fisch}. 
This is largely due to the discovery of the quantum 
Hall effect (QHE) in a 3D or a multilayer system\cite{Stomer}, in 
particular, with the finding of interesting surface states. According 
to the scaling theory of localization\cite{Lee}, all states in a 
two-dimensional (2D) system are localized if only a scalar random 
potential is present. However, in the presence of a strong perpendicular 
magnetic field, where the time reversal symmetry is broken, extended 
states appear in the centers of disorder-broadened Landau bands and 
give rise to the integer QHE. On the other hand, a 3D system with a 
scalar random potential can have a mobility edge that separates extended 
states from localized ones. It is an interesting question to understand 
how the transport phase diagram evolves as a system changes from 3D 
to 2D by weakening the inter-layer coupling in one direction. 

In a multilayer or 3D system with an external magnetic field 
perpendicular to the layers, each energy band will split into several 
Landau subbands. Unlike 2D case, these Landau subbands overlap
with each other because of the third dimension. There may be two 
possible scenarios for the phase diagram of extended and localized 
states in the weak disorder limit. In the case of 
a strong inter-layer coupling, one will expect that there are only 
two mobility edges separating the extended states from the localized 
states as shown in Fig. 1(a), where we sketch the density of
states for a complete band. The system can undergo a 
conventional metal-insulator transition if the Fermi energy of the 
system crosses one of 
the mobility edges. In the metallic phase, electron transport 
perpendicular to layers, or parallel to the magnetic field, is through 
the bulk extended states. In the absence of an inter-layer coupling, 
the system consists of a number of 2D systems. In a strong 
perpendicular magnetic field, it is well known that there are two 
mobility edges for each of the Landau subbands, and the QHE is expected. 
It is natural to expect that a weak coupling can broaden the 
extended state regime, but the localized-extended-localized phase 
diagram may not change in a weak disorder limit
for each of the Landau subbands as shown in 
Fig. 1(b). This is a very interesting scenario with three possible 
phases. The system is an insulator when the Fermi energy is at 
the bottom of the first Landau subband and in the localized 
regime. On the other hand, the system is a normal metal when 
the Fermi energy is inside the extended state region. Finally, the 
system is a 3D quantized Hall conductor\cite{Chalker,Balents} 
when the Fermi energy is in the localized regime between two Landau 
subbands. 
Changing the Fermi energy, the system will undergo 
insulator-metal and metal-quantized Hall conductor transitions. 
In this paper, we study the phase diagram of a weakly 
coupled multilayer system. We show that the second scenario does 
indeed occur for a system with weak disorder. For strong disorder,
there is mobility edge separating extended and localized states,
similar to the zero field case. 

A natural way to address the issue is to study the Anderson 
localization model with a nearest neighbor hopping on a cubic 
lattice. We assume that an external 
magnetic field is along the z-axis. We consider the case of a  
strong magnetic field with 1/3 of a flux quantum per plaquette. 
The hopping coefficients in the x and the y directions are the same, 
and their amplitude is chosen as the energy unit. 
The hopping coefficient along the z-direction is one tenth of 
that in the x or the y direction, in order to describe the case of 
a weak coupling between layers. Therefore, the Hamiltonian of 
this system can be written as:
\begin{equation}
\label{hamiltonia}
{\cal H} = \sum_{\vec{r} } w_{\vec{r}}c_{\vec{r}}^{+} c_{\vec{r}} 
+\sum_{<\vec{r};\vec{\delta}>}t_{\vec{r},\vec{r}+\vec{\delta}}c_{
\vec{r}}^{+}c_{\vec{r}+\vec{\delta}}+c.c
\end{equation}
where $\vec{r}+\vec{\delta}$ labels the nearest neighbors of site 
$\vec{r}$. The disorder potential is modeled by the random on-site 
white-noise potential $w_{\vec{r}}$ ranging from $-W/2$ to $W/2$.
The hopping coefficients are
\begin{equation}
t_{\vec{r},\vec{r}+\vec{\delta}}=\left\{\matrix{
e^{\mp2\pi \alpha yi}, &\vec{\delta}\in\{\pm \hat{\bf{e}}_x\}
\cr 1, &\vec{\delta}\in\{\pm \hat{\bf{e}}_y\} 
\cr 0.1, &\vec{\delta} \in\{\pm \hat{\bf{e}}_z\}
}\right.
\end{equation}
describing a system with a uniform magnetic field $B$ in the 
z-direction with the Landau gauge $\vec{A}=(-By,0,{0})$. The only 
effect of the magnetic field is on the hopping coefficients through 
the Peierls phase\cite{peier} $\alpha=2B/hc$ (we choose lattice 
constant to be one). Without disorder ($W=0$), the energy spectrum can be 
obtained analytically and consists of three bands. The bands are symmetric 
about the center of the second band, which is zero, with the first 
Landau subband below $E=0$ and the third above $E=0$. In the 
presence of disorder, the problem has to be solved numerically.
The localization property of the system is still symmetric 
about $E=0$ even with disorder.

We consider the model on a very long bar geometry of cross section 
$M\times M$ in the $zx$-plane. The bar is along the y-direction. 
Periodic boundary conditions are applied in the x- and z-directions.
Using a standard iteration algorithm, we can calculate the localization 
length $\lambda_M$ at a finite size $M$\cite{wlchan,xwl}. The localization 
length $\lambda_M$ in the critical region is assumed to obey the 
one-parameter scaling law\cite{wlchan,xwl}
\begin{equation}
\label{xivsg}
\Lambda = \frac{\lambda_M}{M} = f\left(\frac{\xi ( \vartheta )}{M} \right),
\end{equation}
where $\xi ( \vartheta )$ is the localization length in 3-D system and $\vartheta$
can be either the randomness of the on-site energy W or the eigen-energy of the 
Hamiltonian. $\xi$ can be found by a least squares fitting method with 
either fixed E or fixed W. The localization length diverges near a 
localized-extended (metal-insulator) transition point as
\begin{equation}
\label{3Dlo}
\xi ( \vartheta )\propto |\vartheta - \vartheta_c|^{-\nu}.
\end{equation}
The transition points $E_c$ or $W_c$ can be determined from the localization 
length $\xi$ from the scaling relation (\ref{3Dlo}). 
In our numerical calculation, we choose the sample length to
be over $10^5$ so that the self-averaging effect automatically takes
care of the ensemble statistical fluctuations.

We now discuss our numerical results. Fig. 2 is the ratio of
finite localization length $\lambda _M$ to the system width $M$
versus energy $E$ with the strength of disorder $W=2$.
The Hamiltonian (\ref{hamiltonia}) is symmetric about E=0, thus, 
one needs to plot the curves for $E>0$ only. 
Different curves are for different system sizes ($\circ: M=8$,
$\ast : M=10$, $\triangle: M=12$, $+: M=14$). All curves 
cross at three points of $E\sim 1.06; \ 1.66; \ 3.22$. Those crossing 
points indicate that there exist three delocalized-localized 
transitions. Near the transition region, the normalized localization 
length $\Lambda = \frac{\lambda_M}{M}$ of finite systems of 
$M=8,10,12,14$ can be described by two nice scaling curves. 
The inset of Fig. 2 is the two-branch scaling curve near the 
mobility edge $E_c=3.22$. 
The transition points can also be obtained through curve-fitting
and we found, again, that $E_c=1.06; 1.66; 3.22$, in good agreement with the 
crossing values. The critical exponent for the localization 
length is found to be about $\nu=1.57\pm 0.10$, consistent with 
previous results\cite{MK,mack,bulk,kramer}.

The results of Fig. 2 show indeed that there is a localized state region, 
$1.06<E<1.66$, between the second and the third Landau 
subbands, as far as the transport vertical to the magnetic
field direction is concerned. Other approaches such as the random network 
approximation\cite{Chalker,Balents} showed that there is a so-called
edge-state sheath in this region. The edge-state sheath is extended 
along the system surface, and electrons can move around through the 
sheath even at zero temperature. Therefore, the system in the region 
behaves like a metal in the sense that the conductance of the system 
along the magnetic field is finite at zero 
temperature\cite{Chalker,Balents,Druist}. However, because the edge-state 
sheath exists near the surface of a sample, the conductance along
the magnetic field depends on the perimeter of the sample rather than 
on the area of the sample cross section as a normal metal 
does\cite{Chalker,Balents,Druist}. Like the 2D edge states give rise to the 
quantum Hall effect, this edge-state sheath makes the Hall conductance 
quantized in 3D. Thus, the system in this region is also called 
a quantized Hall conductor\cite{Chalker,Balents}.
The existence of the edge-state sheath may also be understood from 
that of its 2D counterpart of edge states. It has been shown that an 
extended state can only be destroyed through scattering involving
states in different Landau subbands\cite{lxn}. When the disorder is 
too weak to induce significant mixing among states of different 
Landau subbands, the extended states are preserved in the form of 
edge states. The above argument is also consistent with the 
conservation of the Chern number\cite{lxn}.

We have carried out numerical calculations for many points on the 
$W - E_{c}$ plane for the multilayer system. Presented in Fig. 3 is 
the resulting phase diagram. $E=0$ corresponds to the center 
of the middle band, and, again, the localization properties of the 
first band and the third band are the same (the system is symmetric 
about $E=0$) so we need only to consider $E>0$. There are several 
interesting features about the phase diagram. 1) At weak disorder
($W<4$), there are two mobility edges to separate localized and 
extended states in each of the Landau subbands. Correspondingly, 
there are three delocalized-localized transition points in the 
region of $E>0$. The first transition point at small $E$ is one of 
the two mobility edges for the middle band. 
\noindent The other two are the two 
mobility edges of the third Landau subband. The first transition 
point is the one from normal metal to the quantum Hall conductor.
The second transition point corresponds to the one from the 
quantum-Hall conductor phase to the normal metallic phase. The last 
one at larger $E$ (band edge) describes the transition from the normal 
metallic phase to an insulating phase. Thus, at a weak disorder, there 
is a 3D quantum Hall regime between the first and the second 
transition points. This regime decreases with the disorder. Varying 
the Fermi energy by increasing the electron density, the system 
can enter consecutively into insulator, normal metal, 3D quantum 
Hall conductor, and normal metal again. 
2) The two inner mobility edges of the phase diagram meet at a point 
around $W=4$ to close the pocket of the 3D quantum Hall effect 
regime. At intermediate disorder ($4<W<10.5$), the disorder 
causes the three Landau subbands to couple completely together and 
to form a large energy band. There is only one localized-delocalized 
transition corresponding to the conventional metal-insulator 
transition. 3) At strong disorder ($W>10.5$), all states are 
localized, and the system can only be in an insulating phase. It is 
interesting to notice that the disorder makes an electron state more 
localized, and also broadens the Landau subbands. Below a certain 
value of disorder, the broadening effect dominates over the 
localization effect, and the phase boundary shifts into the localized 
state regime as the disorder increases. Above a certain disorder 
strength, the localization effect dominates, and the extended state 
region decreases with disorder, in agreement with previous 
results\cite{MK,mack,bulk,kramer}.

Before we end this paper, we would like to mention a recent study by 
Drose {\it et. al.}\cite{kramer} who investigated an isotropic Anderson 
model in a strong magnetic field of 1/3 flux quanta per plaquette. They 
did not observe the quantum Hall conductor phase. Their phase diagram
contains only the outermost boundary of Fig. 3 of the current work. This 
is due to the strong layer-layer coupling in their
work which mixes different Landau subbands such that a quantum Hall 
conductor phase cannot exist in the presence 
of disorder. 

In summary, we have demonstrated that there is a region of 3D quantum Hall 
effect in a multilayer system with weak inter-layer coupling at weak 
disorder. For large disorder, the physics is similar to that for the zero field case,
namely, with a mobility edge separating extended and localized states.

We thank Dongzi Liu and Qian Niu for many helpful discussions and 
comments.
X.R. Wang is supported by UGC, Hong Kong, through RGC/DAG grant.
X.C. Xie is supported by DOE under contract number DE-FG03-98ER45687.

 \begin{figure}
 \vspace{2mm}
  \vbox to 14.0cm {\vss\hbox to 6.0cm
  {\hss\
    {\includegraphics{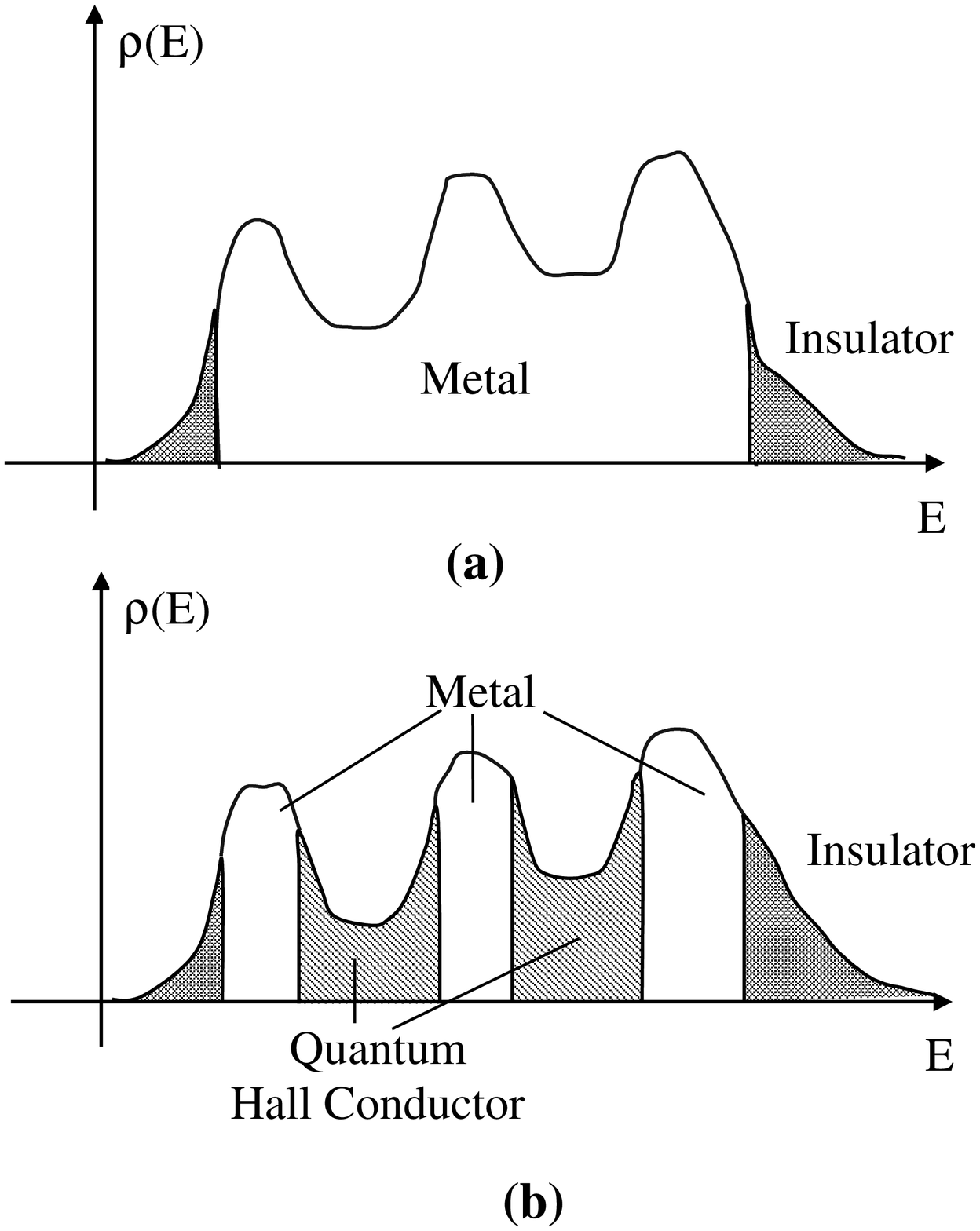}
    }
   \hss}
  } 
\vspace{2mm}
\caption{\label{fig1} Two possible scenarios for a multilayer system in 
a strong magnetic field perpendicular to the layers. a) There is only one
transition separating localized states from extended states for all Landau 
subbands. b) There are mobility edges for each of the Landau subbands, 
separating localized states from extended states. When the Fermi energy is 
between two Landau subbands but inside the localized states regime, the system
is a quantum Hall conductor.}
\end{figure}

\begin{figure}
\vspace{1mm}
 \vbox to 6.0cm {\vss\hbox to 7.0cm 
 {\hss\ 
   {\includegraphics{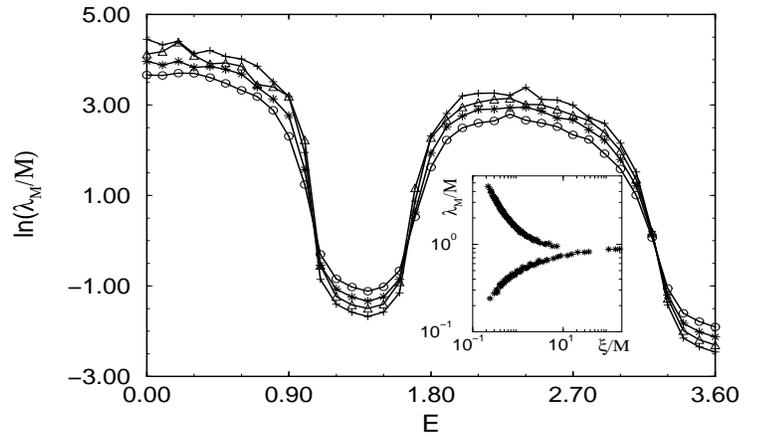} 
   } 
  \hss} 
 }  
\vspace{0.5mm}
\caption{\label{fig2}  $\lambda_{M}/M$ as function of energy $E$
with magnitude of random on-site potential $W=2$. for
$\circ: M=8$, $\ast: M=10$, $\triangle: M=12$,
and $+: M=14$.
Inset: scaling function around the third crossing point $E_c=3.22$.}
\end{figure}
 
\begin{figure}
\vspace{2mm}
 \vbox to 6.0cm {\vss\hbox to 7.0cm 
 {\hss\
   {\includegraphics{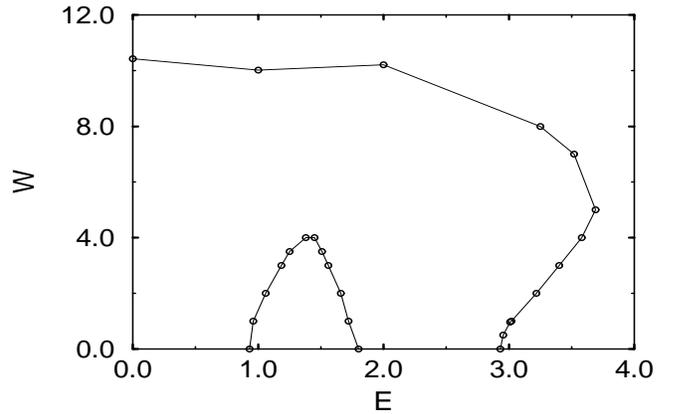}
   } 
  \hss} 
 }  
\vspace{1mm}
\caption{\label{fig3} Phase diagram on the $E_c - W$ plane. The States inside
phase boundary are extended while those outside are localized. The systems
in the middle pocket are in the quantum Hall conductor phase.}
\end{figure}

\end{document}